\documentclass[twocolumn,10pt,twoside]{IEEEtran}
\normalsize

\ifCLASSINFOpdf
\else
\fi

\usepackage{amsfonts}
\usepackage{amsmath}
\usepackage{graphicx}
\usepackage{epstopdf}
\usepackage{color}
\usepackage{multicol}
\usepackage{amssymb}
\usepackage{subfigure}
\usepackage{bm}
\usepackage{latexsym}
\usepackage{stfloats}
\usepackage{enumerate}

\usepackage{algorithm}
\usepackage{listings}
\usepackage{algpseudocode}
\usepackage{cases}

\begin{document}
%
\title{Near-Optimal Hybrid Processing for Massive MIMO Systems via Matrix Decomposition}

%
%
%

\author{
Weiheng~Ni, Xiaodai~Dong, and~Wu-Sheng~Lu
\thanks{
W. Ni, X. Dong and W.-S. Lu are with the Department of Electrical and Computer Engineering, University
of Victoria, Victoria, BC V8W 3P6, Canada. (email: nweiheng@uvic.ca, xdong@ece.uvic.ca,  wslu@ece.uvic.ca).
}
}

\maketitle
\newtheorem{theorem}{Theorem}[section]
\newtheorem{lemma}[theorem]{\textbf{Lemma}}
\newtheorem{proposition}[theorem]{Proposition}
\newtheorem{remark}[theorem]{\textbf{Remark}}
\newtheorem{corollary}[theorem]{Corollary}

\begin{abstract}
For the practical implementation of massive multiple-input multiple-output (MIMO) systems, the hybrid processing (precoding/combining) structure is promising to reduce the high cost rendered by large number of RF chains of the traditional processing structure. The hybrid processing is performed through low-dimensional digital baseband processing combined with analog RF processing enabled by phase shifters. We propose to design hybrid RF and baseband precoders/combiners for multi-stream transmission in point-to-point massive MIMO systems, by directly decomposing the pre-designed unconstrained digital precoder/combiner of a large dimension. The constant amplitude constraint of analog RF processing results in the matrix decomposition problem non-convex. Based on an alternate optimization technique, the non-convex matrix decomposition problem can be decoupled into a series of convex sub-problems and effectively solved by restricting the phase increment of each entry in the RF precoder/combiner within a small vicinity of its preceding iterate. A singular value decomposition based technique is proposed to secure an initial point sufficiently close to the global solution of the original non-convex problem. Through simulation, the convergence of the alternate optimization for such a matrix decomposition based hybrid processing (MD-HP) scheme is examined, and the performance of the MD-HP scheme is demonstrated to be near-optimal. 
\end{abstract}


%
\IEEEpeerreviewmaketitle

\section{Introduction}\label{sec:intro}
Massive multiple-input multiple-output (MIMO) is potentially one of the key technologies to achieve high capacity performance in the next generation of mobile cellular systems \cite{andrews2014what5g}-\cite{larsson2014massive}. In the limit of an infinite number of antennas, the massive MIMO propagation channel becomes quasi-static, where the effects of uncorrelated noise and fast fading vanish, and such favorable characteristics enables arbitrarily small energy per transmitted bit \cite{marzetta2010noncoop_mimo}. Prominently, in massive multiuser MIMO systems simple linear processing schemes, such as zero-forcing (ZF) and linear minimum mean-square error (MMSE), are shown to approach the optimal capacity performance achieved by the dirty paper coding in the downlink communication \cite{erez2005dpc}. The spectral efficiency performance of massive MIMO systems with several linear processing schemes, including ZF, MMSE and maximum-ratio combining (MRC), with perfect or imperfect channel state information (CSI) has been analyzed in \cite{hien2013efficiency}.

For practical implementation of massive MIMO systems, the number of antennas required for large antenna array gains, typically in the order of a hundred or more, is determined by examining the convergence properties over the antenna number  \cite{smith2014convergence}. However, to exploit such a large antenna array in massive MIMO systems, the amplitudes and phases of the complex transmit symbols are traditionally modified at the baseband, and then upconverted to the passband around the carrier frequency after passing through radio frequency (RF) chains (performing the analog radiowave/digital baseband conversion, signal mixing, power amplifying). In this setting, all outputs of the RF chains are connected to the antenna elements, which means that the number of the RF chains must be exactly equal to the number of antenna elements.  Under the circumstances the fabrication cost and energy consumption of such a massive MIMO system become unbearable due to the tremendous number of RF chains \cite{liang2014hybrid}. 

To deal with the aforementioned problem, smaller number of RF chains are used in the large scale MIMO systems, where cost-effective variable phase shifters can be employed to handle the mismatch between the number of RF chains and of antennas \cite{zhang2005antselect}-\cite{liu2014rfonly}, where high-dimensional analog RF (phase only) processing is enabled by using phase shifters while digital baseband processing is performed in a very low dimension. In \cite{zhang2005antselect}, both diversity and multiplexing transmissions of MIMO communications are addressed with a limited number of RF chains. In \cite{love2003equalgain} and \cite{zheng2007uniform}, analog RF precoding is presented to achieve full diversity order and near-optimal beamforming performance. Reference \cite{liu2014rfonly} applies phase-only RF precoding to massive MIMO systems to maximize the data rate of users based on a bi-convex approximation approach. Especially, the utilization of small wavelengths of millimeter wave (mmWave) makes it possible to build a large antenna array in a compact region. This hybrid baseband and RF processing (transmit precoding/receive combining) scheme is found particularly suitable for mmWave MIMO communications as it effectively reduces the excessive cost of RF chains \cite{heath2012beamsteer}-\cite{sayeed2013beamspace}. Herein, hybrid processing is designed to capture ``dominant'' paths in point-to-point (P2P) mmWave channels by choosing RF control phases from array response vectors \cite{heath2012beamsteer}, \cite{heath2014sparse}. On the other hand, hybrid processing in multiuser mmWave systems is investigated in \cite{liang2014hybrid}, \cite{kim2013multibeam}-\cite{sayeed2013beamspace}, where analog RF processing aims to obtain large antenna gains, while baseband processing performs in low-dimensional equivalent channels.

More often than not, CSI is the prerequisite to perform any processing at transmitter and receiver, whether it is a type of unconstrained high-dimensional baseband processing for the traditional design with one antenna element coupled with one dedicated RF chain or it is a type of hybrid processing. In \cite{andrew2014closedloop}, training sequences and closed-loop sounding vectors are designed to estimate a massive multiple-input single-output (MISO) channel through the alignment of transmit beamformer with true channel direction. In \cite{nquyen2013chestimation}, a compressive sensing (CS) based low-rank approximation problem for estimating massive MIMO channel matrix is formulated, and is solved by semidefinite programming. Considering the massive MIMO channels with limited scattering feature (especially when mmWave channels are involved), the parameters of paths, such as the angles of departure (AoDs), angles of arrival (AoAs) and the corresponding path loss are estimated by designing beamforming codebook so as to obtain the pathloss of all paths whose AoDs/AoAs are spatially quantized in the entire angular domain \cite{dinesh2012csadaption}, \cite{ahmed2014channel}, while \cite{ahmed2014channel} performs the beamforming in a hybrid processing setting. 

In this paper, we propose to design hybrid RF and baseband precoders/combiners for multi-stream transmission in P2P massive MIMO systems by directly decomposing the pre-designed unconstrained digital precoder/combiner of a large dimension. This is an approach that has not been attempted in the literature. In our design, the analog RF precoder/combiner are constrained by the nature of the phase shifters so that the amplitudes of all entries of the RF precoder/combiner matrices remain constant.  Starting with an optimal unconstrained precoder built on a set of right singular vectors (associated with the largest singular values) of the channel matrix, our hybrid precoders are designed by minimizing the Frobenius norm of the matrix of difference between the unconstrained precoding matrix and products of the hybird RF and baseband precoding matrices, subject to the aforementioned constraints on the RF precoder. Technically, solving this matrix decomposition problem is rather challenging because it is a highly nonconvex constrained problem involving a fairly large number of design parameters. Here we present an alternate optimization technique to approach the solution in that the hybrid precoders are alternately optimized in a relaxed setting so as to ensure all sub-problems involved are convex. We stress that the convex relaxation technique utilized here includes not only properly grouping design parameters for alternate optimization, but also restricting the phase increment of each entry in the RF precoder to within a small vicinity of its preceding iterate. Under these circumstances, it is critical to start the proposed decomposition algorithm with a suitable initial point that is sufficiently close to the global solution of the original non-convex matrix decomposition problem. To this end a singular-value-decomposition (SVD) based technique is proposed to secure a satisfactory initial point that with high probability allows our decomposition algorithm to yield near-optimal hybrid precoders. Concerning the hybrid combiners design, a linear MMSE combiner is selected as the unconstrained reference matrix for matrix decomposition, and the hybrid RF and baseband combiners can be obtained in the same way as the hybrid precoder design. 
 
We remark that the matrix decomposition based hybrid processing design scheme, termed as MD-HP, is suited for hybrid processing design over any general massive MIMO channels as long as the channel matrix is assumed to be known. Simulations are presented to examine the convergence of the alternate optimization for the MD-HP scheme and to demonstrate the near-optimal performance of the MD-HP scheme by comparing it to the optimal unconstrained baseband processing based on SVD technique.

\section{System Model}\label{sec:sys_model}
In this section, we introduce the hybrid processing structure for P2P massive MIMO systems and the channel models considered in this paper.
\subsection{System Model}
\begin{figure}[htbp]
    \centering
    \includegraphics[width=0.35\textwidth]{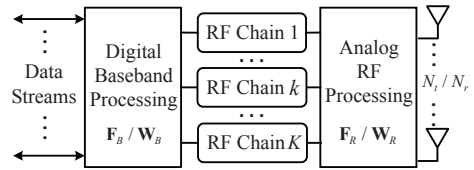}
    \caption{System model of the transceiver with the hybrid processing structure}
    \label{fig:system_model}
\end{figure}

We consider a communication scenario from a transmitter with $N_t$ antennas and $M_t$ RF chains to a receiver equipped with $N_r$ antennas and $M_r$ RF chains, where $N_s$ data streams are supported. The system model of the transceiver is shown in Fig. \ref{fig:system_model}. To ensure   effectiveness of the communication driven by the limited number of RF chains, the number of the communication streams is constrained to be bounded by $N_s \leq M_t \leq N_t$ for the transmitter and by $N_s \leq M_r \leq N_r$ for the receiver.

The transmitted symbols are processed by a baseband precoder $\mathbf{F}_B$ of dimension $M_t \times N_s$, then up-converted to the RF domain through the $M_t$ RF chains before being precoded by an RF precoder $\mathbf{F}_R$ of dimension $N_t \times M_t$. Note that the baseband precoder $\mathbf{F}_B$ enables both amplitude and phase modifications, while only phase changes can be realized by $\mathbf{F}_R$ as it is implemented by using analog phase shifters. We normalize each entry of $\mathbf{F}_R$ to satisfy $|\mathbf{F}_R^{(i,j)}| = \frac{1}{N_t}$, where $|(\cdot)^{(i,j)}|$ denotes the amplitude of the $(i, j)$-th element of $(\cdot)$. Furthermore, to meet the constraint on total transmit power, $\mathbf{F}_B$ is normalized to satisfy $||\mathbf{F}_R\mathbf{F}_B||_F^2 = N_s$, where $||\cdot||_F$ denotes the Frobenius norm \cite{golub1996matrix}.

We assume a narrowband flat fading channel model and the received signal is given by
\begin{equation}
    \mathbf{y} = \mathbf{H} \mathbf{F}_R\mathbf{F}_B \mathbf{s} + \mathbf{n},
\end{equation}
where $\mathbf{y} \in \mathbb{C}^{N_r \times 1}$ is the received signal vector, $\mathbf{s} \in \mathbb{C}^{N_s \times 1}$ is the signal vector such that $\mathbb{E}[\mathbf{ss}^H] = \frac{P}{N_s}\mathbf{I}_{N_s}$ where $(\cdot)^H$ denotes conjugate transpose, $\mathbb{E}[\cdot]$ denotes expectation, $\mathbf{I}_{N_s}$ is the $N_s \times N_s$ identity matrix and $P$ is the average transmit power. $\mathbf{H} \in \mathbb{C}^{N_r \times N_t}$ is the channel matrix, normalized as $\mathbb{E}[||{\mathbf{H}}_F^2||] = N_tN_r$, and $\mathbf{n}$ is the vector of i.i.d. $\mathcal{CN}(0, \sigma^2)$ addictive complex Gaussian noise. To perform the precoding and combining, we assume the channel is known at both the transmitter and the receiver, thus the processed received signal after combining is given by
\begin{equation}
    \tilde{\mathbf{y}} = \mathbf{W}_B^H\mathbf{W}_R^H \mathbf{H} \mathbf{F}_R\mathbf{F}_B \mathbf{s} + \mathbf{W}_B^H\mathbf{W}_F^H\mathbf{n},
\end{equation}
where $\mathbf{W}_F$ is the $N_r \times M_r$ RF combining matrix and $\mathbf{W}_B$ is the $M_r \times N_s$ baseband combining matrix. Since $\mathbf{W}_F$ is also implemented by the analog phase shifters, all elements of $\mathbf{W}_F$ are constrained to have constant amplitude such that $|\mathbf{W}_B^{(i,j)}| = \frac{1}{N_r}$. If Gaussian inputs are employed at the transmitter, the long-term average spectral efficiency achieved shall be
\begin{equation}
\begin{aligned}
    R(\mathbf{F}_R, \mathbf{F}_B, \mathbf{W}_R, \mathbf{W}_B) = \log_2 \left( \left| \mathbf{I}_{N_s} + \frac{P}{N_s}\mathbf{R}_n^{-1} \tilde{\mathbf{H}}\tilde{\mathbf{H}}^H \right| \right),
\end{aligned} 
\end{equation}
where $\mathbf{R}_n = \sigma^2\mathbf{W}_B^H\mathbf{W}_F^H\mathbf{W}_F\mathbf{W}_B$ is the covariance matrix of the noise and $\tilde{\mathbf{H}} =  \mathbf{W}_B^H\mathbf{W}_F^H \mathbf{H} \mathbf{F}_R\mathbf{F}_B$.

\subsection{Channel Model}
In this paper, we seek to find optimal hybrid precoders $(\mathbf{F}_R$, $\mathbf{F}_B)$ as well as hybrid combiners $(\mathbf{W}_R$, $\mathbf{W}_B)$ based on a general channel matrix $\mathbf{H}$. To measure the performance of our MD-HP scheme, we examine two types of channel models in simulation studies to be presented in Section \ref{sec:simulation}, namely
\begin{itemize}
    \item[1)] Large Rayleigh fading channel $\mathbf{H}_{rl}$ with i.i.d. $\mathcal{CN}(0, 1)$ entries;
    \item[2)] Limited scattering mmWave channel $\mathbf{H}_{mmw}$.
\end{itemize}

We remark that several hybrid processing schemes for mmWave communications have been studied, where a large antenna array is implemented to combat high free-space pathloss and reflection loss \cite{heath2012beamsteer}-\cite{sayeed2013beamspace}. Thus the mmWave channel model $\mathbf{H}_{mmw}$ is an appropriate instance for comparing the performance of the proposed scheme with related recent work in the literature. Because of the limited (sparse) scattering characteristic of a mmWave channel, we decide to introduce a clustered mmWave channel model to characterize its key features \cite{spencer2000anglech}. The mmWave channel $\mathbf{H}_{mmw}$ is assumed to be the sum of all propagation paths that are scattered in $N_c$ clusters with each cluster contributing $N_p$ paths. Under these circumstances, the normalized channel matrix can be expressed as
\begin{equation}\label{eq:cl_ch_model}
    \mathbf{H}_{mmw} = \sqrt{\frac{N_tN_r}{N_cN_p}}\sum_{i = 1}^{N_c}\sum_{l=1}^{N_p}\alpha_{il} \mathbf{a}_r(\theta_{il}) \mathbf{a}_t(\phi_{il})^H,
\end{equation}
where $\alpha_{il}$ is the complex gain of the $i$-th path in the $l$-th cluster, which follows $\mathcal{CN}(0, 1)$\footnote{The power gain of the channel matrix is normalized such that $\mathbb{E}[||{\mathbf{H}_{mmw}}_F^2||] = N_tN_r$}. For the $(i, l)$-th path, $\theta_{il}$ and $\phi_{il}$ are the azimuth angles of arrival/departure (AoA/AoD), while $\mathbf{a}_r(\theta_{il})$ and $\mathbf{a}_t(\phi_{il})$ are the receive and transmit array response vectors at the azimuth angles of $\theta_{il}$ and $\phi_{il}$ respectively, and the elevation dimension is ignored\footnote{Only 2D beamforming is considered in this mmWave channel model.}. Within the $i$-th cluster, $\theta_{il}$ and $\phi_{il}$ have the uniformly-distributed mean values of $\theta_i$ and $\phi_i$ respectively, while the lower and upper bounds of the uniform distribution for $\theta_i$ and $\phi_i$ can be defined as $[\theta_{\min}, \theta_{\max}]$ and $[\phi_{\min}, \phi_{\max}]$. The angular spreads (standard deviations) of $\theta_{il}$ and $\phi_{il}$ among all clusters are assumed to be constant, denoted as $\sigma_\theta$ and $\sigma_\phi$. According to \cite{heath2014sparse}, we use truncated Laplacian distribution to generate all the  $\theta_{il}$'s and $\phi_{il}$'s based on the above parameters. 

As for the array response vectors $\mathbf{a}_r(\theta_{il})$ and $\mathbf{a}_t(\phi_{il})$, we choose uniform linear arrays (ULAs) in our simulations, while the precoding scheme to be developed in Section \ref{sec:precoding} can directly be applied to arbitrary antenna arrays. For an $N$-element ULA, the array response vector can be given by
\begin{equation}\label{eq:ula}
    \mathbf{a}_{ULA}(\theta) = \frac{1}{\sqrt{N}}\left[ 1, e^{j\frac{2\pi}{\lambda}d\sin(\theta)}, \cdots, e^{j(N-1)\frac{2\pi}{\lambda}d\sin(\theta)} \right]^T,
\end{equation}
where $\lambda$ is the wavelength of the carrier, and $d$ is the distance between any two adjacent antenna elements. The array response vectors at both the transmitter and the receiver can be written in the form of (\ref{eq:ula}).

\section{Hybrid Precoding/Combining Design for A General Massive MIMO Channel}\label{sec:precoding}
The design of hybrid precoders $(\mathbf{F}_R$, $\mathbf{F}_B)$ and combiners $(\mathbf{W}_R$, $\mathbf{W}_B)$ based on a general massive MIMO channel $\mathbf{H}$ may be achieved by formulating a joint transmitter-receiver optimization problem to maximize the spectral efficiency, which is given by
\begin{equation}\label{eq:joint_opt}
\begin{aligned}
    \max~& R(\mathbf{F}_R, \mathbf{F}_B, \mathbf{W}_R, \mathbf{W}_B) \\
    s.t.~& ||\mathbf{F}_R\mathbf{F}_B||_F^2 = N_s, \\
         & \mathbf{F}_R \in \mathcal{F}_R,~\mathbf{W}_R \in \mathcal{W}_R,
\end{aligned}
\end{equation}
where $\mathcal{F}_R (\mathcal{W}_R)$ is the set of matrices with all constant amplitude entries, which is $\frac{1}{\sqrt{N_t}} (\frac{1}{\sqrt{N_r}})$. However, this type of joint optimization problems is often intractable \cite{escalante2011nonconvex}, due to the presence of non-convex constraints $\mathbf{F}_R \in \mathcal{F}_R$ and $\mathbf{W}_R \in \mathcal{W}_R$ that obstruct the regular progress of securing a globally optimal solution. Before gaining an insight into the solution of this joint optimization problem (\ref{eq:joint_opt}), we introduce the optimal unconstrained precoder $\mathbf{F}^\star$ and combiner $\mathbf{W}^\star$ for achieving maximum capacity of a general MIMO channel, based on which a procedure for the design of near-optimal hybrid precoders/combiners is developed. Assume that the channel matrix $\mathbf{H}$ is well-conditioned to transmit $N_s$ data streams, namely, $\mathrm{rank}(\mathbf{H}) \geq N_s$. To obtain the optimal $\mathbf{F}^\star$ and $\mathbf{W}^\star$, we perform the SVD of the channel matrix $\mathbf{H = U\Sigma V}^H$, where $\mathbf{U}$ and $\mathbf{V}$ are $N_r \times N_r$ and  $N_t \times N_t$  unitary matrices, respectively, and $\mathbf{\Sigma}$ is an $N_r \times N_t$ diagonal matrix with singular values on its diagonal in descendant order. 
Without incorporating the waterfilling power allocation, the optimal unconstrained precoder and combiner are given by 
\begin{equation}\label{eq:uv_partition}
\mathbf{F}^\star = \mathbf{V}_1,~~\mathbf{W}^\star = \mathbf{U}_1,
\end{equation}
where $\mathbf{V}_1$ and $\mathbf{U}_1$ are constructed with the first $N_s$ columns of $\mathbf{V}$ and $\mathbf{U}$, respectively, and the corresponding spectral efficiency by using such unconstrained $\mathbf{F}^\star$ and $\mathbf{W}^\star$ is given by
\begin{equation}
    \tilde{R} = \log_2 \left( \left| \mathbf{I}_{N_s} + \frac{\gamma}{N_s} \mathbf{\Sigma}_1^2 \right| \right),
\end{equation}
where $\mathbf{\Sigma}_1$ represents the first partition of dimension $N_s \times N_s$ of $\mathbf{\Sigma}$ by defining that
\begin{equation}
	\mathbf{\Sigma} = \left[ \begin{array}{cc}
	\mathbf{\Sigma}_1 & \mathbf{0} \\
	\mathbf{0} & \mathbf{\Sigma}_2 \\
	\end{array} \right],
\end{equation}
where $\gamma = \frac{P}{\sigma^2}$ is the signal-to-noise ratio (SNR).

Actually, $\tilde{R}$ sets an upper bound for the spectral efficiency $R(\mathbf{F}_R, \mathbf{F}_B, \mathbf{W}_R, \mathbf{W}_B)$ in problem (\ref{eq:joint_opt}) where the ranges of the matrix products $\mathbf{F}_R\mathbf{F}_B$ and $\mathbf{W}_R\mathbf{W}_B$ are respectively the subsets of feasible regions of the unconstrained precoder and combiner, namely, $\mathbb{C}^{N_t \times N_s}$ and $\mathbb{C}^{N_r \times N_s}$. Considering the non-convex nature of the problem (\ref{eq:joint_opt}), it is impractical to insist upon securing its global solution. One apparently viable approach is to construct hybrid precoders $(\mathbf{F}_R, \mathbf{F}_B)$ and combiners $(\mathbf{W}_R, \mathbf{W}_B)$ such that the optimal unconstrained precoder $\mathbf{F}^\star$ and combiner $\mathbf{W}^\star$ can be sufficiently closely approached by $\mathbf{F}_R\mathbf{F}_B$ and $\mathbf{W}_R\mathbf{W}_B$ respectively. In what follows, the design of such hybrid precoders is substantiated via matrix decomposition.

\subsection{Hybrid Precoders Design via Matrix Decomposition}\label{sec:hybrid_precoder}
Given hybrid precoding structure and constraint on RF precoder $\mathbf{F}_R$, there is no guarantee that a pair $(\mathbf{F}_R, \mathbf{F}_B)$ can be found such $\mathbf{F}^\star = \mathbf{F}_R\mathbf{F}_B$ holds exactly. However, by relaxing the strict equality in (\ref{eq:joint_opt}), the matrix decomposition can be accomplished through reformulating the original problem as
\begin{equation}\label{eq:precoder_decomp}
\begin{aligned}
    \min_{\mathbf{F}_R, \mathbf{F}_B}~& ||\mathbf{F}^\star - \mathbf{F}_R\mathbf{F}_B||_F\\
    s.t.~& ||\mathbf{F}_R\mathbf{F}_B||_F^2 = N_s, \\
         & \mathbf{F}_R \in \mathcal{F}_R.
\end{aligned}
\end{equation}


To look closely at the physical implication of this problem re-formulation, recall that our design objective is essentially to approximate $\mathbf{F}^\star$ by the product of hybrid precoding matrices, namely $\mathbf{F}_R\mathbf{F}_B$. Thus a natural question arising at this point is how sensitive the spectral efficiency $R(\mathbf{F}_R, \mathbf{F}_B, \mathbf{W}_R, \mathbf{W}_B)$ to any deviation of $\mathbf{F}_R\mathbf{F}_B$ from $\mathbf{F}^\star$, because small residue $||\mathbf{F}^\star - \mathbf{F}_R\mathbf{F}_B||_F$ at a solution of problem (\ref{eq:precoder_decomp}) is inevitable and this residue would divert the optimal unconstrained combiner $\mathbf{W}^\star$ away from the SVD-based solution $\mathbf{U}_1$. 

Bearing the analysis above on mind, we begin the design of hybrid precoders by assuming that the $N_r$--dimensional minimum distance decoding can be performed at the receiver, which implies that the achieved spectral efficiency is equivalent to the mutual information over the MIMO channel when Gaussian inputs are used, which is given by
\begin{equation}\label{eq:mutual_info}
    \mathcal{I}(\mathbf{F}_R, \mathbf{F}_B) = \log_2 \left( \left| \mathbf{I}_{N_s} + \frac{\gamma}{N_s}\mathbf{H} \mathbf{F}_R\mathbf{F}_B \mathbf{F}_B^H\mathbf{F}_R^H \mathbf{H}^H \right| \right).
\end{equation}
Next, we obtain the hybrid precoders by maximizing the mutual information in (\ref{eq:mutual_info}). The problem of mutual information maximization problem has been investigated in \cite{heath2014sparse}, where the mutual information is approximated as
\begin{equation}\label{eq:mutual_info_approx}
\begin{aligned}
    \mathcal{I}&(\mathbf{F}_R, \mathbf{F}_B) \\
    & \approx \log_2 \left( \left| \mathbf{I}_{N_s} + \frac{\gamma}{N_s} \mathbf{\Sigma}_1^2 \right| \right) - N_s + ||\mathbf{V}_1^H\mathbf{F}_R\mathbf{F}_B||_F^2,
\end{aligned} 
\end{equation}
and $\max\mathcal{I}(\mathbf{F}_R, \mathbf{F}_B) \approx \max||\mathbf{V}_1^H\mathbf{F}_R\mathbf{F}_B||_F^2$ is approximately equivalent to minimizing $||\mathbf{F}^\star - \mathbf{F}_R\mathbf{F}_B||_F$. Consequently, designing $(\mathbf{F}_R, \mathbf{F}_B)$ so as to maximize the mutual information over the massive MIMO channel can be accomplished by solving the matrix decomposition problem (\ref{eq:precoder_decomp}). Once the hybrid precoders $(\mathbf{F}_R, \mathbf{F}_B)$ are optimized, we can proceed to design the hybrid combiners $(\mathbf{W}_R, \mathbf{W}_B)$ to maximally increase the system's spectral efficiency.

The second constraint in (\ref{eq:precoder_decomp}) requiring that the entries of $\mathbf{F}_R$ have constant amplitude $\frac{1}{\sqrt{N_t}}$ is evidently non-convex, which the use of efficient convex optimization algorithms and makes it extremely challenging to secure a globally optimal solution. Under the circumstance, our design searches for a near-optimal solution so that the spectral efficiency achieved by the obtained hybrid precoders (as well as the hybrid combiners) is comparable with the upper bound $\tilde{R}$. The design method described below has three main ingredients: it employs an alternate optimization strategy that separates the two sets of design parameters in a natural manner; a local convexification technique ensures that each sub-problem be solved in a convex setting; and the use of a carefully chosen initial point that facilitates the alternate iterates to converge to a satisfactory design. 

Alternate minimization is an iterative procedure with each iteration be carried out in two steps. In each of these steps one set of design parameters are fixed while the objective function is minimized with respect to the other set of parameters and the role of design parameters alternates as the design step switches. For the design problem at hand, naturally the components in $\mathbf{F}_R$ and those in $\mathbf{F}_B$ are the two parameter sets, the alternate minimization is performed as follows: 1) solve problem (\ref{eq:precoder_decomp}) with respect to $\mathbf{F}_B$ with $\mathbf{F}_R$ given; and 2) solve problem (\ref{eq:precoder_decomp}) with respect to $\mathbf{F}_R$ with $\mathbf{F}_B$ given.  

We begin by examining a simplified version of problem (\ref{eq:precoder_decomp}) by temporarily removing the normalization constraint $||\mathbf{F}_R\mathbf{F}_B||_F^2 = N_s$, which leads (\ref{eq:precoder_decomp}) to
\begin{equation}\label{eq:precoder_decomp_simplified}
\begin{aligned}
    \min_{\mathbf{F}_R, \mathbf{F}_B}~& ||\mathbf{F}^\star - \mathbf{F}_R\mathbf{F}_B||_F\\
    s.t.~& \mathbf{F}_R \in \mathcal{F}_R.
\end{aligned}
\end{equation}
Denote the hybrid precoders at the $k$-th iteration by $(\mathbf{F}_R^{(k)}, \mathbf{F}_B^{(k)})$, and assume the initial $\mathbf{F}_R^{(0)}$ is given. We update $\mathbf{F}_B^{(k)}$ by solving the unconstrained convex problem $\min_{\mathbf{F}_B} ||\mathbf{F}^\star - \mathbf{F}_R^{(k)}\mathbf{F}_B^{(k)}||_F$ whose closed-form solution is given by 
\begin{equation}
    \mathbf{F}_B^{(k)} = ({\mathbf{F}_R^{(k)}}^H\mathbf{F}_R^{(k)})^{-1}{\mathbf{F}_R^{(k)}}^H\mathbf{F}^\star,~k=0, 1, 2, \cdots.
\end{equation}
In turn, we update the RF precoder to $\mathbf{F}_R^{(k+1)}$ by solving the non-convex problem below while $\mathbf{F}_B^{(k)}$ is given as a constant matrix:
\begin{equation}\label{eq:update_fr}
\begin{aligned}
    \min_{\mathbf{F}_R^{(k+1)}}~& ||\mathbf{F}^\star - \mathbf{F}_R^{(k+1)}\mathbf{F}_B^{(k)}||_F\\
    s.t.~& \mathbf{F}_R^{(k+1)} \in \mathcal{F}_R.
\end{aligned}
\end{equation}
To deal with the nonconvex constraint $\mathbf{F}_R^{(k+1)} \in \mathcal{F}_R$ in (\ref{eq:update_fr}), we update $\mathbf{F}_R^{(k)}$ with a local search in a small vicinity of $\mathbf{F}_R^{(k)}$. Denote the phase of the $(m, n)$-th entry of $\mathbf{F}_R^{(k)}$ as $\phi_{m,n}^{(k)}$, $\mathbf{F}_R^{(k)}$ can be represented as as $\frac{1}{\sqrt{N_t}}\{e^{j\phi_{m,n}^{(k)}}\},~m = 1,\cdots,N_t,~n = 1,\cdots,M_t$. To characterize the relation between $\mathbf{F}_R^{(k+1)}$ and $\mathbf{F}_R^{(k)}$, we write $\mathbf{F}_R^{(k+1)}$ as 
\begin{equation}\label{eq:fr_kplus1}
    \mathbf{F}_R^{(k+1)} = \frac{1}{\sqrt{N_t}}\{e^{j\phi_{m,n}^{(k+1)}}\} = \frac{1}{\sqrt{N_t}}\{e^{j(\phi_{m,n}^{(k)} + \delta_{m,n}^{(k)})}\},
\end{equation}
where $\delta_{m,n}^{(k)}$ is the phase increment of the $(m, n)$-th entry of $\mathbf{F}_R^{(k)}$. Note that the approximation $e^{j\delta_{m,n}^{(k)}} \approx 1 + j\delta_{m,n}^{(k)}$ holds as long as $|\delta_{m,n}^{(k)}|$ is sufficiently small, e.g. $|\delta_{m,n}^{(k)}| \leq 0.1$. Based on Taylor's expansion, therefore, we have
\begin{equation}\label{eq:fr_kpkus1_approx}
\begin{aligned}
    \mathbf{F}_R^{(k+1)} &\approx \frac{1}{\sqrt{N_t}} \{(1 + j\delta_{m,n}^{(k)}) e^{j\phi_{m,n}^{(k)}}\} \\
    &= \mathbf{F}_R^{(k)} + \frac{j}{\sqrt{N_t}}\{\delta_{m,n}^{(k)} \cdot e^{j\phi_{m,n}^{(k)}}\}\\
    &= \mathbf{F}_R^{(k)} + \{\delta_{m,n}^{(k)}\} \circ \frac{j}{\sqrt{N_t}}\{e^{j\phi_{m,n}^{(k)}}\},
\end{aligned} 
\end{equation}
where $\{\delta_{m,n}^{(k)}\}$ is the matrix whose $(m,n)$-th entry is $\delta_{m,n}^{(k)}$ and ``$\circ$" denotes the Hadamard product (entrywise product). It follows that the problem in (\ref{eq:update_fr}) for seeking $\mathbf{F}_R^{(k+1)}$ can be reformulated as an optimization problem with respect to $\{\delta_{m,n}^{(k)}\}$ as 
\begin{equation}\label{eq:update_fr_delta}
\begin{aligned}
    &\min_{\{\delta_{m,n}^{(k)}\}}~ \left|\left|\mathbf{F}^\star - \left(\mathbf{F}_R^{(k)} + \{\delta_{m,n}^{(k)}\} \circ \frac{j}{\sqrt{N_t}}\{e^{j\phi_{m,n}^{(k)}}\}\right)\mathbf{F}_B^{(k)}\right|\right|_F \\
    \Leftrightarrow &\min_{\{\delta_{m,n}^{(k)}\}}~ \left|\left|\mathbf{Q}^{(k)} - \left(\{\delta_{m,n}^{(k)}\} \circ \frac{j}{\sqrt{N_t}}\{e^{j\phi_{m,n}^{(k)}}\}\right)\mathbf{F}_B^{(k)}\right|\right|_F^2,
\end{aligned} 
\end{equation}
where $\mathbf{Q}^{(k)} = \mathbf{F}^\star - \mathbf{F}_R^{(k)}\mathbf{F}_B^{(k)}$. 
We remark that problem (\ref{eq:update_fr_delta}) has a convex quadratic objective function, and the constant amplitude constraint $\mathbf{F}_R^{(k+1)} \in \mathcal{F}_R$ has been into account because $\mathbf{F}_R^{(k+1)}$ here assumes the form of $\frac{1}{\sqrt{N_t}}\{e^{j(\phi_{m,n}^{(k+1)})}\}$. However, the above formulation is based on the approximation $e^{j\delta_{m,n}^{(k)}} \approx 1 + j\delta_{m,n}^{(k)}$, hence it is valid only if $|\delta_{m,n}^{(k)}|$ is sufficiently small. Therefore, linear constraints on the smallness of $|\delta_{m,n}^{(k)}|$ need to be imposed, thus problem (\ref{eq:update_fr_delta}) is modified to
\begin{equation}\label{eq:update_fr_small_delta}
\begin{aligned}
    \min_{\{\delta_{m,n}^{(k)}\}}~ &\left|\left|\mathbf{Q}^{(k)} - \left(\{\delta_{m,n}^{(k)}\} \circ \frac{j}{\sqrt{N_t}}\{e^{j\phi_{m,n}^{(k)}}\}\right)\mathbf{F}_B^{(k)}\right|\right|_F^2 \\
    s.t.~ &|\delta_{m,n}^{(k)}| \leq \bar{\delta}^{(k)}, \forall m,n,
\end{aligned} 
\end{equation}
where $\bar{\delta}^{(k)} > 0$ is sufficiently small such that $e^{j\delta_{m,n}^{(k)}} \approx 1 + j\delta_{m,n}^{(k)}$ holds. Problem (\ref{eq:update_fr_small_delta}) is a convex quadratic programming (QP) problem whose unique global solution can be calculated efficiently \cite{boyd2004opt}. Once the solution $\{\delta_{m,n}^{(k)}\}$ is obtained, the $\mathbf{F}_R^{(k+1)}$ can be updated by (\ref{eq:fr_kplus1}). 

There are several issues that remain to be addressed. These include defining an error measure to be used in the algorithm's stopping criterion and elsewhere; selection of a good initial point to start the algorithm; adaptive thresholding for phase increments $\delta_{m,n}^{(k)}$ and derivation of an explicit formulation for problem (\ref{eq:update_fr_small_delta}); and a treatment of  the constraint $||\mathbf{F}_R\mathbf{F}_B||_F^2 = N_s$ in problem (\ref{eq:precoder_decomp}).

\subsubsection{An Error Measure}
The relative distance between $\mathbf{F}^\star$ and $\mathbf{F}_R^{(k)}\mathbf{F}_B^{(k)}$, namely $\varepsilon_k = \frac{||\mathbf{F}^\star - \mathbf{F}_R^{(k)}\mathbf{F}_B^{(k)}||_F}{||\mathbf{F}^\star||_F}$, will be used as an error measure. In the proposed algorithm, alternate iterations continue until $|\varepsilon _k - \varepsilon _{k-1}|$ falls below a prescribed convergence tolerance $\bar{\varepsilon}$, and when this occurs, the last iterate $(\mathbf{F}_R^{(k)}, \mathbf{F}_B^{(k)})$ is taken to be a solution of problem (\ref{eq:precoder_decomp_simplified}). 

\subsubsection{Adaptive Thresholding for Phase Increments $\delta_{m,n}^{(k)}$}\label{sec:dynamic_delta}
The constraints on the magnitude of phase increment in (\ref{eq:update_fr_small_delta}) limit $\mathbf{F}_R^{(k+1)}$ to within a small neighborhood of $\mathbf{F}_R^{(k)}$ that usually affects the algorithm's convergence rate. This is however less problematic for (\ref{eq:update_fr_small_delta}) because the effective range for each phase parameter in the RF precoder is limited to $[0, 2\pi)$. In addition, the issue can be addressed by making the upper bound (threshold) in (\ref{eq:update_fr_small_delta}) adaptive to the current error measure so as to improve the algorithm's convergence rate. The adaptation of threshold $\bar{\delta}^{(k)}$ is performed as follows:
\begin{itemize}
    \item[1)] set $\bar{\delta}^{(k+1)}$ slightly larger than $\bar{\delta}^{(k)}$ if $\varepsilon_k$ is far greater than $\bar{\varepsilon}$, and $\varepsilon_k < \varepsilon_{k-1}$ holds;
    \item[2)] set a smaller $\bar{\delta}^{(k+1)}$ than $\bar{\delta}^{(k)}$ if $\varepsilon_k$ is close to $\bar{\varepsilon}$, or $\varepsilon_k \geq \varepsilon_{k-1}$ holds.
\end{itemize}
Scenario 1) allows a larger phase increment while the algorithm converges in the right direction ($\varepsilon_k$ is decreasing), while scenario 2) reduces $\bar{\delta}^{(k)}$ to a smaller $\bar{\delta}^{(k+1)}$ when $\varepsilon_k$ increases due to that the previous large phase increment has made the approximation $e^{j\delta_{m,n}^{(k)}} \approx 1 + j\delta_{m,n}^{(k)}$ invalid, or when $\varepsilon_k$ is close to the required $\bar{\varepsilon}$ suggesting that higher precision is required. In Section \ref{sec:simulation} we shall come back to this matter again in terms of specific adjustments on $\bar{\delta}^{(k)}$.

\subsubsection{Re-formulation of Problem (\ref{eq:update_fr_small_delta})}
Another issue concerning problem (\ref{eq:update_fr_small_delta}) is that its formulation in terms of Hadamard product is not suited for many convex-optimization solvers that require standard and explicit formulations. Denote the $p$-th row of $\mathbf{Q}^{(k)}$ by $\mathbf{q}_p^{(k)}$, we can write the objective function in (\ref{eq:update_fr_small_delta}) as
\begin{equation}\label{eq:objfunc_sum_fnorm}
\begin{aligned}
    &\sum_{p=1}^{N_t}\left|\left|\mathbf{q}_p^{(k)} - \left[ \frac{j\delta_{p,1}^{(k)}}{\sqrt{N_t}}e^{j\phi_{p,1}^{(k)}}, \cdots, \frac{j\delta_{p,N_t}^{(k)}}{\sqrt{N_t}}e^{j\phi_{p,N_t}^{(k)}}\right]\mathbf{F}_B^{(k)}\right|\right|_2^2 \\
    =& \sum_{p=1}^{N_t}\left|\left|\mathbf{q}_p^{(k)} - \mathbf{\Delta}_p^{(k)} \mathbf{G}_p^{(k)}\right|\right|_2^2,
\end{aligned} 
\end{equation}
where $\mathbf{\Delta}_p^{(k)} = [\delta_{p,1}^{(k)}, \delta_{p,2}^{(k)}, \cdots, \delta_{p,N_t}^{(k)}]$ and $\mathbf{G}_p^{(k)} = \frac{j}{\sqrt{N_t}}\mathbf{diag}\left(e^{j\phi_{p,1}^{(k)}}, \cdots, e^{j\phi_{p,N_t}^{(k)}}\right)\mathbf{F}_B^{(k)}$, hence 
\begin{equation}\label{eq:minofsum_sumofmin}
\begin{aligned}
\min_{\{\delta_{m,n}^{(k)}\}} \sum_{p=1}^{N_t}\left|\left|\mathbf{q}_p^{(k)} - \mathbf{\Delta}_p^{(k)} \mathbf{G}_p^{(k)}\right|\right|_2^2 \\
= \sum_{p=1}^{N_t}\min_{\mathbf{\Delta}_p^{(k)}}\left|\left|\mathbf{q}_p^{(k)} - \mathbf{\Delta}_p^{(k)} \mathbf{G}_p^{(k)}\right|\right|_2^2.
\end{aligned}
\end{equation}
It follows that problem (\ref{eq:update_fr_small_delta}) can be solved by separately solving $N_t$ sub-problems
\begin{equation}\label{eq:delta_subpro}
\begin{aligned}
    \min_{\mathbf{\Delta}_p^{(k)}}~&\left|\left|\mathbf{q}_p^{(k)} - \mathbf{\Delta}_p^{(k)} \mathbf{G}_p^{(k)}\right|\right|_2^2 \\
    s.t.~&|\delta_{p,n}^{(k)}| \leq \bar{\delta}^{(k)}, \forall n,
\end{aligned}
\end{equation}
for $p = 1,2,\cdots, N_t$. Note that each problem in (\ref{eq:delta_subpro}) is an explicitly formulated convex QP problem to which efficient interior-point algorithms apply \cite{boyd2004opt}.

\subsubsection{Choosing an Initial Point}\label{sec:initial_point}
Choosing an appropriate initial point to start the proposed algorithm is of critical importance because the original problem (\ref{eq:precoder_decomp}) is a non-convex problem which typically possesses multiple local minimizers. As far as gradient-based optimization algorithms are concerned, the likelihood of capturing global minimizer or a good local minimizer is known to be highly dependent on how close the initial point to the desired solution.

Note that the objective function in (\ref{eq:precoder_decomp}), namely $||\mathbf{F}^\star - \mathbf{F}_R\mathbf{F}_B||_F$, measure the difference between the optimal unconstrained RF precoder $\mathbf{F}^\star$ and an actual decomposition $\mathbf{F}_R\mathbf{F}_B$ in the feasible region. If we temporarily neglect the constant amplitude constraint on $\mathbf{F}_R$, the perfect decomposition of $\mathbf{F}^\star$ can be performed through SVD decomposition $\mathbf{F}^\star = \mathbf{U}_F\mathbf{\Sigma}_F\mathbf{V}_F^H$. This motivates us to construct an initial point based on the SVD of $\mathbf{F}^\star$. As $\mathbf{F}^\star$ comes from the first $M_t$ right singular vectors of the channel matrix $\mathbf{H}$, $\mathbf{F}^\star$ has the full column rank, which means all $N_s$ entries along the diagonal of $\mathbf{\Sigma}_F$ are non-zero. Note that $\mathbf{U}_F\mathbf{\Sigma}_F$ is an $N_t \times N_s$ matrix with full column rank, $\mathbf{V}_F^H$ is an $N_s \times N_s$ matrix and $\mathbf{F}_R$ consists of $N_t^{RF}$ columns. To construct an initial point that conforms to the dimensions of $\mathbf{F}_R$ and $\mathbf{F}_B$, we generate an $N_t \times (M_t-N_S)$ matrix $\hat{\mathbf{F}}_R$ where the amplitude of each entry is equal to $\frac{1}{\sqrt{N_t}}$ and the phase of each entry obeys a uniform distribution over $[0, 2\pi)$. In this way, a decomposition of $\mathbf{F}^\star$ is found to be
\begin{equation}\label{eq:fopt_decomp}
    \mathbf{F}^\star = [\mathbf{U}_F\mathbf{\Sigma}_F~~\hat{\mathbf{F}}_R] 
    \left[ \begin{array}{c}
	\mathbf{V}_F^H  \\
	\mathbf{0} \\
	\end{array} \right],
\end{equation}
and $(\mathbf{F}_R = [\mathbf{U}_F\mathbf{\Sigma}_F~~\hat{\mathbf{F}}_R],~\mathbf{F}_B = [\mathbf{V}_F~~\mathbf{0}]^H)$ is exactly a global solution for $\min||\mathbf{F}^\star - \mathbf{F}_R\mathbf{F}_B||$ when no constraints are imposed. We stress that $\mathbf{F}_R = [\mathbf{U}_F\mathbf{\Sigma}_F~~\hat{\mathbf{F}}_R]$ is infeasible when the constant amplitude constraint of $\mathbf{F}_R$ is imposed. Nevertheless, we can select a feasible initial point $\mathbf{F}_R^{(0)}$ that is close to the above $[\mathbf{U}_F\mathbf{\Sigma}_F~~\hat{\mathbf{F}}_R]$ by modifying the first partition $\mathbf{U}_F\mathbf{\Sigma}_F$ as follows:
\begin{itemize}
    \item[1)] retaining the phases of all entries in $\mathbf{U}_F\mathbf{\Sigma}_F$;
    \item[2)] enforcing the amplitudes of all entries in $\mathbf{U}_F\mathbf{\Sigma}_F$ into $\frac{1}{\sqrt{N_t}}$ to make $\mathbf{F}_R^{(0)}$ feasible.
\end{itemize}
Since the modified $\mathbf{U}_F\mathbf{\Sigma}_F$ still incorporates the information of the phases in decomposition ($\ref{eq:fopt_decomp}$), it is intuitively clear that the $\mathbf{F}_R^{(0)}$ generated above is reasonably near the global solution of problem (\ref{eq:precoder_decomp_simplified}) and for this reason we shall chose it as the initial point for the proposed algorithm.

Finally, the constraint $||\mathbf{F}_R\mathbf{F}_B||_F^2 = N_s$ in the original matrix decomposition problem (\ref{eq:precoder_decomp}) is treated by performing a normalization step where $\mathbf{F}_B$ is multiplied by $\frac{\sqrt{N_s}}{||\mathbf{F}_R\mathbf{F}_B||_F}$. The normalization assures that the transmission power remains consistent after precoding. A step-by-step summary of the hybrid precoder design is given below as Algorithm \ref{algo:alter_opt_precoder}. 
\begin{algorithm}[htbp]
    \caption{The Hybrid Precoders Design via Matrix Decomposition based on Alternating Optimization}
    \label{algo:alter_opt_precoder}
    \begin{algorithmic}[1]
    \Require $\mathbf{F}^\star$, $\mathbf{F}_R^{(0)}$ 
    \State $\mathbf{F}_B^{(0)} = ({\mathbf{F}_R^{(0)}}^H\mathbf{F}_R^{(0)})^{-1}{\mathbf{F}_R^{(0)}}^H\mathbf{F}^\star$
    \State $\varepsilon_0 = \frac{||\mathbf{F}^\star - \mathbf{F}_R^{(0)}\mathbf{F}_B^{(0)}||_F}{||\mathbf{F}^\star||_F}$, $\varepsilon_{-1} = \infty$ 
    \State $k = 0$
    \While {$|\varepsilon_k - \varepsilon_{k-1}| \leq \bar{\varepsilon} $}
        \State $k = k + 1$ 
        \State obtain $\mathbf{F}_R^{(k)}$ by solving (\ref{eq:update_fr}) 
		\State $\mathbf{F}_B^{(k)} = ({\mathbf{F}_R^{(k)}}^H\mathbf{F}_R^{(k)})^{-1}{\mathbf{F}_R^{(k)}}^H\mathbf{F}^\star$
        \State $\varepsilon_k = \frac{||\mathbf{F}^\star - \mathbf{F}_R^{(k)}\mathbf{F}_B^{(k)}||_F}{||\mathbf{F}^\star||_F}$
    \EndWhile 
    \State $\mathbf{F}_B = \frac{\sqrt{N_s}\mathbf{F}_B}{||\mathbf{F}_R\mathbf{F}_B||_F}$ \\
    \Return $\mathbf{F}_R$, $\mathbf{F}_B$
    \end{algorithmic}
\end{algorithm}

\subsection{Hybrid Combiners Design}
The hybrid precoders are designed under the assumption that the $N_r$--dimensional minimum distance decoding can be performed at the receiver. However, such a decoding scheme is difficult to implement in practice due to its high complexity. In this paper, we employ linear combining at the receiver. As we are aware, if the hybrid precoders would be equivalent to the unconstrained optimal precoder $\mathbf{F}^\star = \mathbf{V}_1$, the optimal unconstrained combiner $\mathbf{W}^\star$ would be $\mathbf{U}_1$. However the error $||\mathbf{F}^\star - \mathbf{F}_R\mathbf{F}_B||_F$ can never be absolutely zero, hence $\mathbf{U}_1$ deviate from the optimal unconstrained combiner $\mathbf{W}^\star$. The linear MMSE combiner $\mathbf{W}_{MMSE}$ will achieve the maximum spectral efficiency when only linear combination is performed before detection and only 1-dimensional detection is allowed for each data stream. The unconstrained linear MMSE combiner is given in \cite{kailath2000mmse} as
\begin{equation}\label{eq:mmse_combiner}
\begin{aligned}
    \mathbf{W}^\star &= \mathbf{W}_{MMSE} \\
    &= \arg\min_{\mathbf{W}}~\mathbb{E}\left[ ||\mathbf{s} - \mathbf{Wy}||_2 \right]\\
    &= \frac{\sqrt{P}}{N_s}
        \left( \frac{P}{N_s}\mathbf{H}\mathbf{F}_R\mathbf{F}_B\mathbf{F}_B^H\mathbf{F}_R^H\mathbf{H}^H + \sigma^2\mathbf{I}_{N_r} \right)^{-1} \mathbf{H}\mathbf{F}_R\mathbf{F}_B.
\end{aligned}
\end{equation}
Once $\mathbf{W}^\star$ is obtained, the alternate optimization method presented above can be directly applied to decompose $\mathbf{W}^\star$ into hybrid combiners $\mathbf{W}_R$ and $\mathbf{W}_B$, which leads to the problem
\begin{equation}\label{eq:combiner_decomp}
\begin{aligned}
    \min_{\mathbf{W}_R, \mathbf{W}_B}~& ||\mathbf{W}^\star - \mathbf{W}_R\mathbf{W}_B||_F\\
    s.t.~& \mathbf{W}_R \in \mathcal{W}_R.
\end{aligned}
\end{equation}

We will evaluate the performance of the proposed hybrid processing scheme through simulations in Section \ref{sec:simulation}.

\subsection{Approach To Waterfilling Spectral Efficiency}
To reach a capacity-achieving processing scheme, the waterfilling power allocation should be applied to the precoder. In this case, the optimal unconstrained precoder and combiner in Section \ref{sec:precoding} are updated to $\mathbf{F}^\star = \mathbf{V}_1\mathbf{\Gamma}$ and $\mathbf{W}^\star = \mathbf{U}_1$ respectively, where $\mathbf{\Gamma}$ is a diagonal matrix that performs the waterfilling power allocation. The precoder so produced can directly be decomposed through Algorithm \ref{algo:alter_opt_precoder}. However, there may be cases where no power is allocated to some data streams corresponding to the lowest singular values of $\mathbf{H}$, especially when the SNR is small. In other words, we may end up with $\mathbf{F}^* = [\mathbf{F}^\prime, \mathbf{0}]$, where $\mathbf{F}^\prime$ is the non-zero columns of $\mathbf{F}^\star = \mathbf{V}_1\mathbf{\Gamma}$ after waterfilling power allocation. In this case, we can apply the MD-HP scheme to the $\mathbf{F}^\prime$ part first, $\mathbf{F}^\prime = \mathbf{F}_{R}\mathbf{F}_{B}^\prime$. And then the whole decomposition for $\mathbf{F}^*$ is given by $\mathbf{F}^* = [\mathbf{F}^\prime, \mathbf{0}] = [\mathbf{F}_{R}\mathbf{F}_{B}^\prime, \mathbf{0}] = \mathbf{F}_{R}[\mathbf{F}_{B}^\prime, \mathbf{0}] = \mathbf{F}_{R}\mathbf{F}_{B}$. In this way, the zero-power allocation part is realized through the baseband precoding rather than the phase shift in the RF domain.

\subsection{Quantized RF Phase Control}
It is difficult to assign arbitrary value to the phase of each entry in the RF precoder $\mathbf{F}_R$ or combiner $\mathbf{W}_R$ is difficult to be set to be an arbitrary value due to the limited precision in the practical implementation. To address the problem here, we also introduce the quantized phase implementation of $\mathbf{F}_R$ and $\mathbf{W}_R$. Assume the phase of each entry in $\mathbf{F}_R$ and $\mathbf{W}_R$ can be quantized up to $L$ bits of precision by choosing the closet neighbor based on the shortest Euclidean distance, which is given by
\begin{equation}\label{eq:quantize}
    \phi = \frac{2\pi\bar{n}}{2^L},
\end{equation}
where $\bar{n} = \arg \min_{n \in \{0, \cdots, 2^L-1\}}\left| \phi - \frac{2\pi n}{2^L} \right|$.

\section{Simulation Results}\label{sec:simulation}
In this section, we report the results of the simulations conducted, where the convergence of the proposed matrix decomposition method based on alternate optimization are examined and the performance of the proposed MD-HP scheme are evaluated.

\subsection{Convergence Properties of Algorithm \ref{algo:alter_opt_precoder}}
Before we apply Algorithm \ref{algo:alter_opt_precoder} to design the hybrid precoders and combiners, it is necessary to examine whether it will converge to a level where the error $\varepsilon_k$ is acceptably small, this is because the original optimization problem (\ref{eq:precoder_decomp}) to be solved is non-convex and there is no guarantee that Algorithm \ref{algo:alter_opt_precoder} will certainly result in a satisfactory matrix decomposition. 

We took a $256 \times 64$ MIMO system as example, and set $N_s = 4$, $M_t = 6$. An i.i.d Rayleigh fading channel matrix $\mathbf{H}_{rl}$ with each entry obeying $\mathcal{CN}(0, 1)$ was randomly generated. From Section \ref{sec:hybrid_precoder}, the optimal unconstrained precoder $\mathbf{F}^\star$ was obtained by selecting the first $N_s$ right singular vectors based on the SVD decomposition on $\mathbf{H}_{rl}$. An initial RF precoder $\mathbf{F}_R^{(0)}$ was chosen by following the technique described in Section \ref{sec:initial_point}. The threshold was set to $\bar{\varepsilon}=10^{-5}$ and the first  phase increment threshold was set to $\bar{\delta}^{(1)}=0.1$. In the simulations, two options for $\bar{\delta}^{(k)}$ were examined:
\begin{itemize}
    \item[1)] $\bar{\delta}^{(k)} = 0.1,~\forall k$;
    \item[2)] $\bar{\delta}^{(k)} = \left\{\begin{array}{c}
    1.25\cdot\bar{\delta}^{(k-1)},~\mathrm{when}~|\varepsilon_{k-1}-\varepsilon_{k-2}| > 100\cdot\bar{\varepsilon}  \\
    0.8\cdot\bar{\delta}^{(k-1)},~\mathrm{when}~|\varepsilon_{k-1}-\varepsilon_{k-2}| \leq 100\cdot\bar{\varepsilon} \\
    \end{array}\right.$.
\end{itemize}
For option 2) with adaptive phase increment threshold, the adjustment of $\bar{\delta}^{(k)}$ depends on how close the previous two error indicators are. When the difference of the previous error indicators is smaller than $100\cdot\bar{\varepsilon}$, which means Algorithm \ref{algo:alter_opt_precoder} is going to converge, $\bar{\delta}^{(k)}$ should be reduced to enhance the precision of the solution by guaranteeing the effectiveness of the approximation $e^{j\delta_{m,n}^{(k)}} \approx 1 + j\delta_{m,n}^{(k)}$. Otherwise, $\bar{\delta}^{(k)}$ can be augmented to accelerate the algorithm by enlarging the feasible region of (\ref{eq:update_fr_small_delta}). Moreover, we need to decrease $\bar{\delta}^{(k)}$ whenever $\varepsilon_{k-1} > \varepsilon_{k-2}$ which means the previous $\bar{\delta}^{(k-1)}$ is too large to guarantee $e^{j\delta_{m,n}^{(k)}} \approx 1 + j\delta_{m,n}^{(k)}$. We restricted $\bar{\delta}^{(k)} \in [0.1, 0.5]$ by clamping $\bar{\delta}^{(k)}$ to $0.1 (0.5)$ when it was smaller (larger) than $0.1 (0.5)$ in case that the feasible region for (\ref{eq:update_fr_small_delta}) was too small or too large\footnote{All parameters given in this section can be revised for other specific cases}. 

\begin{figure}[htbp]
    \centering
    \includegraphics[width=0.43\textwidth]{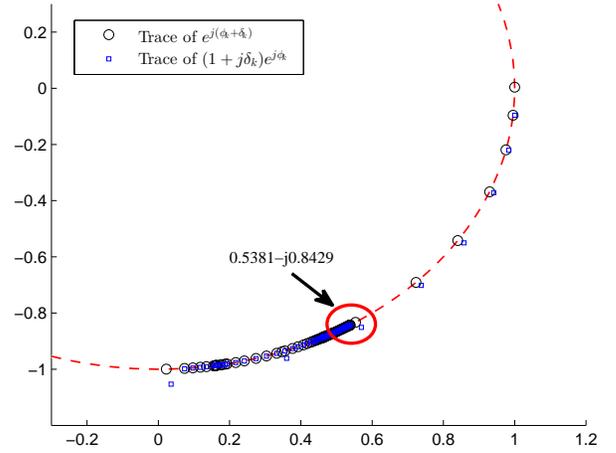}
    \caption{The traces of $e^{j(\phi_{m,n}^{(k)}+\delta_{m,n}^{(k)})}$ and $(1 + j\delta_{m,n}^{(k)})e^{j\phi_{m,n}^{(k)}}$ on the complex plane.}
    \label{fig:phi_trace}
\end{figure}

\begin{figure}[htbp]
    \centering
    \includegraphics[width=0.45\textwidth]{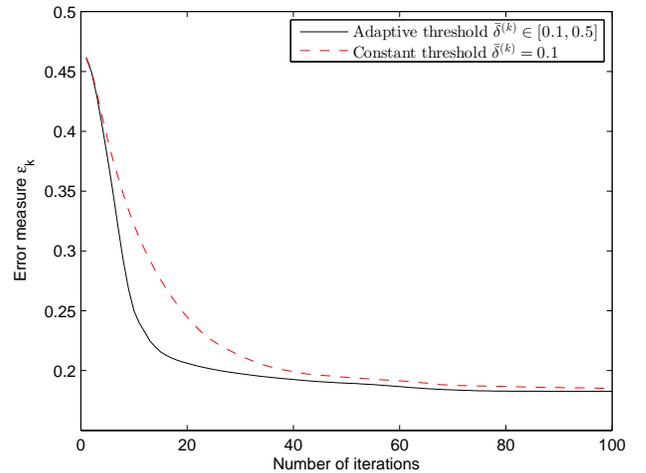}
    \caption{The convergence performance of Algorithm \ref{algo:alter_opt_precoder} when applying the adaptive and constant $\bar{\delta}^{(k)}$ respectively}
    \label{fig:dynamic_delta}
\end{figure}

To examine the effectiveness of approximation $e^{j\phi_{m,n}^{(k+1)}} = e^{j(\phi_{m,n}^{(k)}+\delta_{m,n}^{(k)})} \approx (1 + j\delta_{m,n}^{(k)})e^{j\phi_{m,n}^{(k)}}$, we compare the traces of $e^{j(\phi_{m,n}^{(k)}+\delta_{m,n}^{(k)})}$ and $(1 + j\delta_{m,n}^{(k)})e^{j\phi_{m,n}^{(k)}}$ within 100 iterations in Fig. \ref{fig:phi_trace}, where the red dash line indicates the unit circle on the complex plane. It is observed that the points of the two traces ($m=1, n=5$) update simultaneously and two corresponding points remain very close, which suggests that the iteration $e^{j\phi_{m,n}^{(k+1)}} = e^{j(\phi_{m,n}^{(k)}+\delta_{m,n}^{(k)})}$ may be regarded as a linear operation over $\delta_{m,n}^{(k)}$. By performing adaptive $\bar{\delta}^{(k)}$, $e^{j\phi_{m,n}^{(k)}}$ updates with relatively larger step size at the beginning when the iterate is far from the solution $e^{-j1.0026} \approx 0.5381 - j0.8429$, and then gradually gets close to it. In Fig. \ref{fig:dynamic_delta}, we show how the error measure $\varepsilon_k$ converges to about 0.2 as the number of iterations increases when the adaptive and constant $\bar{\delta}^{(k)}$ are applied respectively. It can be observed that the adaptive threshold $\bar{\delta}^{(k)}$ helps the algorithm converge more quickly because it allows the algorithm to conduct a search over a larger part of the feasible region when the error $\varepsilon_{k}$ is relative small. The above parameters will also be used in the next simulations.

\subsection{Spectral Efficiency Evaluation}
In this part of simulation section, we illustrate the spectral efficiency performance of the proposed MD-HP scheme by comparing it with several other options under large i.i.d. Rayleigh channel and mmWave channel settings respectively. The SNR $\gamma = \frac{P}{\sigma^2}$ range was set to be from -40 dB to 0 dB in all simulations.
\subsubsection{Large i.i.d Rayleigh Fading Channels}
\begin{figure}[htbp]
    \centering
    \includegraphics[width=0.45\textwidth]{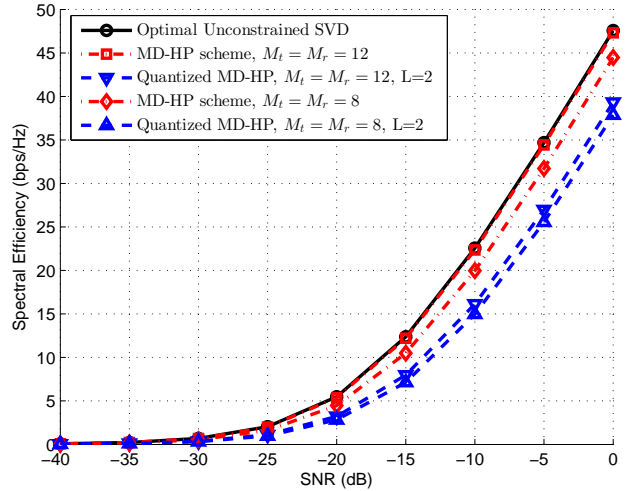}
    \caption{Spectral efficiency achieved by different processing schemes of a $256 \times 64$ massive MIMO system in i.i.d. Rayleigh fading channels where $N_s=8$ data streams are transmitted through 8 and 12 RF chains respectively.}
    \label{fig:rayleigh_m812n8}
\end{figure}

\begin{figure}[htbp]
    \centering
    \includegraphics[width=0.45\textwidth]{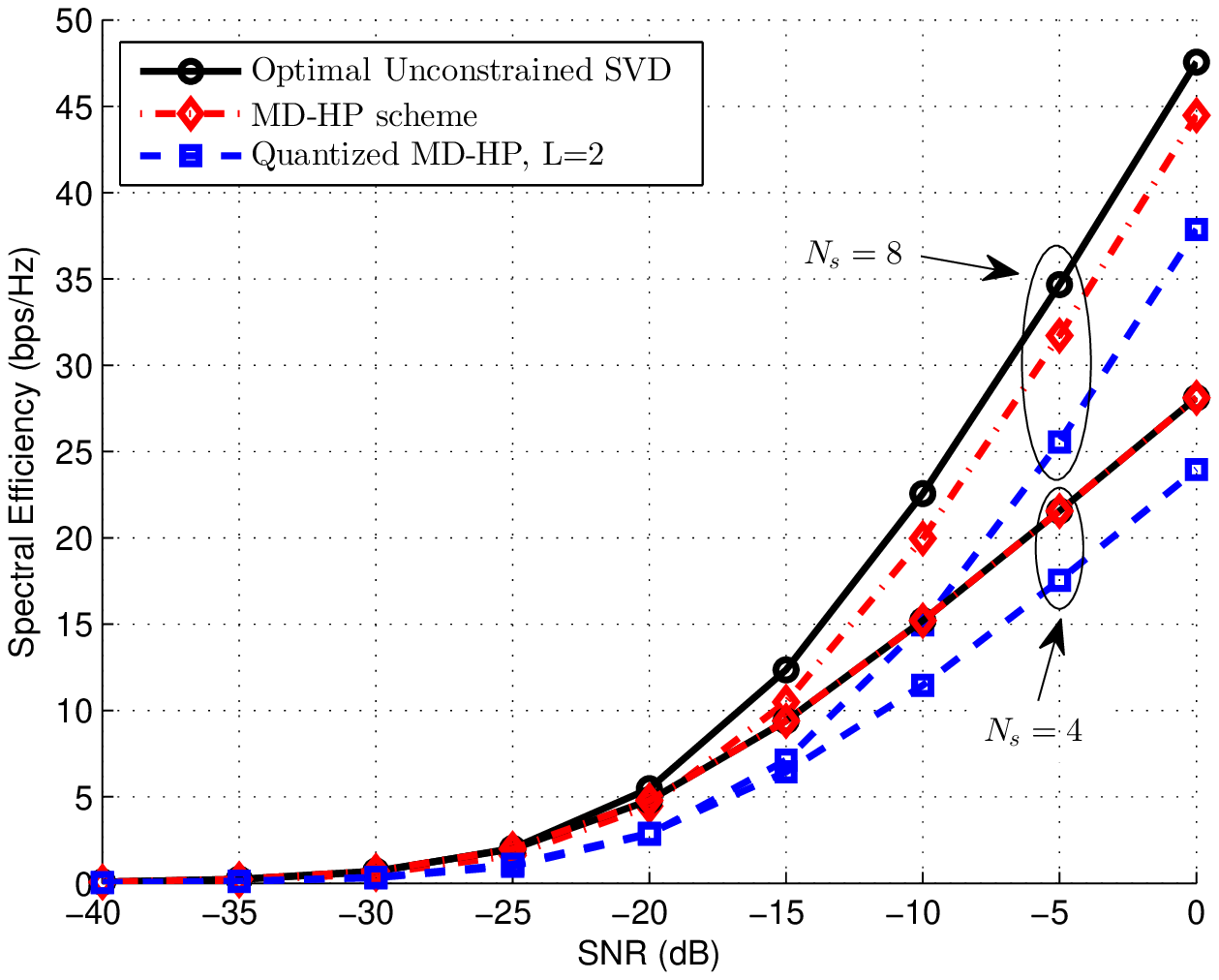}
    \caption{Spectral efficiency achieved by different processing schemes of a $256 \times 64$ massive MIMO system in i.i.d. Rayleigh fading channels where $N_s=4$ and $8$ data streams are transmitted through 8 RF chains respectively.}
    \label{fig:rayleigh_m8n48}
\end{figure}

\begin{figure}[htbp]
    \centering
    \includegraphics[width=0.45\textwidth]{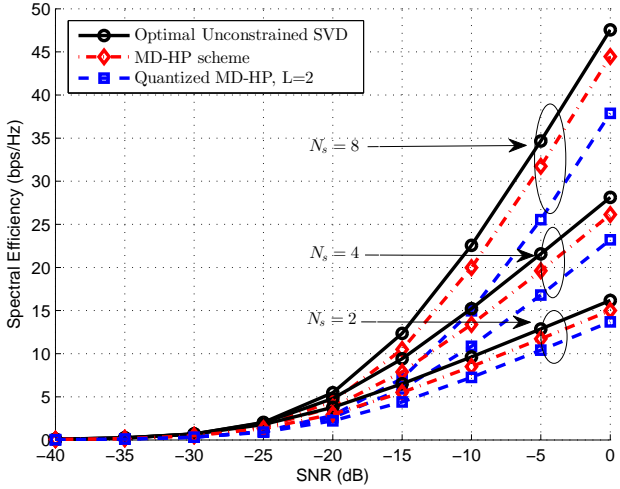}
    \caption{Spectral efficiency achieved by different processing schemes of a $256 \times 64$ massive MIMO system in i.i.d. Rayleigh fading channels where $N_s=2, 4$ and $8$ data streams are transmitted respectively and the numbers of RF chains are set to $M_t = M_r = N_s$.}
    \label{fig:rayleigh_ns248}
\end{figure}

The MD-HP scheme is compared in Fig. \ref{fig:rayleigh_m812n8} against the optimal unconstrained SVD based processing scheme when $N_s = 8$ data streams are transmitted in a $256 \times 64$ massive MIMO system. For the MD-HP scheme, the situations of using 8 and 12 RF chains (along with their quantized versions) are examined. When 12 RF chains are implemented at both the transmitter and receiver, the performance of the MD-HP scheme is near-optimal compared with the optimal unconstrained SVD based scheme. Even though we reduce the number of the RF chains to the number of the data streams, namely, 8 RF chains are employed, the spectral efficiency achieved by the MD-HP scheme slightly decreases by around 3 bps/Hz. As for the heavily quantized versions ($L = 2$ bits with the phase candidates $\{0, \frac{1}{2}\pi, \pi, \frac{3}{2}\pi\}$) corresponding to the 8 and 12 RF chains settings, the spectral efficiency suffers less than 2.5 dB loss (from the view of SNR). Fig. \ref{fig:rayleigh_m8n48} further demonstrates the spectral efficiency performance by also setting the number of transmit data streams to 4 while 8 RF chains are used. Compared with the case of 4 transmit data streams, the performance of the 8 data stream case is evidently improved thanks to the multiplexing gain. Notably, there is a small gap between the MD-HP scheme and the SVD based scheme which can be eliminated by properly increasing the number of RF chains, e.g., double the number of the data streams in the case of $N_s=4$. In addition, the quantized versions ($L=2$) also results in 2.5 dB loss in performance. Under a critical condition that the numbers of RF chains of the transmitter and receiver are set to $M_t = M_r = N_s$, Fig. \ref{fig:rayleigh_ns248} shows the spectral efficiency of the above schemes with $N_s = 2, 4$ and $8$. It is observed that the MD-HP scheme (including the quantized version) consistently remains close to the optimal spectral efficiency as $N_s$ increases, which implies that the MD-HP scheme can probably achieve the near-optimal performance even when a large number of data streams are conducted.

\subsubsection{Large mmWave Channels}
\begin{figure}[htbp]
    \centering
    \includegraphics[width=0.45\textwidth]{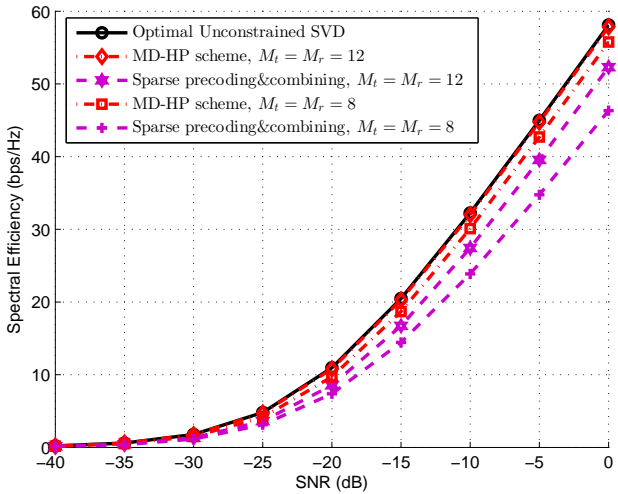}
    \caption{Spectral efficiency achieved by different processing schemes of a $256 \times 64$ massive MIMO system in mmWave channels where $N_s=8$ data streams are transmitted through 8 and 12 RF chains respectively.}
    \label{fig:mmwave_m812n8_256}
\end{figure}

Our proposed MD-HP scheme can also be applied to the large mmWave channels where a certain number of hybrid processing schemes have been studied in the literature. In simulations, the clustered mmWave channel model (\ref{eq:cl_ch_model}) was adopted to characterize its limited scattering feature. Apart from the unconstrained SVD based processing and our MD-HP schemes, we employ the spatially sparse processing \cite{heath2014sparse} which designs the hybrid precoders/combiners by capturing the characteristics of the dominant paths. The propagation model mainly follows the settings in \cite{heath2014sparse}: 1) the mmWave channel incorporates $N_c = 8$ clusters, each of which has $N_p = 10$ paths; 2) the transmitter angle sector is assumed to be $60^\circ$-wide in the azimuth while the receiver with a smaller omni-directional antenna array; 3) the angle spreads of the transmitter and receiver $\sigma_\theta$ and $\sigma_\phi$ are all set to be $7.5^\circ$; 4) the antenna spacing $d$ is equal to half-wavelength.
In Fig. \ref{fig:mmwave_m812n8_256}, the spectral efficiency performance is demonstrated in a $256 \times 64$ mmWave MIMO system, where $N_s = 8$ data streams are transmitted through 8 or 12 RF chains. Our proposed MD-HP scheme apparently outperforms the spatially sparse processing scheme when the same number of RF chains are implemented. Moreover, the MD-HP scheme can even achieve higher spectral efficiency with only 8 RF chains than the spatially sparse processing scheme with 12 RF chains. Particularly, the SVD based processing is sufficiently approached by the MD-HP scheme given 12 RF chains. It is shown that our proposed MD-HP scheme can better capture the characteristics of the mmWave channel than the spatially sparse processing scheme.

\section{Conclusion}
In this paper, we have designed the hybrid RF and baseband precoders/combiners for multi-stream transmission in P2P massive MIMO systems via solving a non-convex matrix decomposition problem. Based on an alternate optimization technique, we have transformed the non-convex matrix decomposition problem into a series of convex sub-problems. Careful handling of the phase increment of each entry in RF precoders and combiners in each iteration and smart choice of an initial point have allowed our algorithm to yield near-optimal solution with high probability. The MD-HP scheme can be applied to any general massive MIMO channels such as i.i.d. Rayleigh fading channels and mmWave channels. By providing enough number of RF chains (e.g., double the number of the transmit data streams), the pre-designed unconstrained digital precoder/combiner of a large dimension can be sufficiently approached and thus the near-optimal performance is achieved. A low quantization level such as 2 bits for phase shifters has been shown to lead to around 2.5 dB loss in performance. We aim to incorporate channel estimation and reduce the time complexity of the MD-HP scheme in the future.

%




\end{document}